\documentclass[%
 reprint,
superscriptaddress,
 amsmath,amssymb,
 aps,
prd,
nofootinbib
]{revtex4-1}

\usepackage{graphicx}
\usepackage{dcolumn}
\usepackage{bm}
\usepackage{hyperref}
\hypersetup{
    colorlinks=true,
    linkcolor=blue,
    filecolor=magenta,      
    urlcolor=blue,
    citecolor=blue
    }
\usepackage[capitalise,nameinlink]{cleveref}
\usepackage[english]{babel}
    
\begin{document}

\preprint{APS/123-QED}

\title{Dense output for highly oscillatory numerical solutions}

\author{F.J. Agocs}
\email{fa325@cam.ac.uk}
\affiliation{Astrophysics Group, Cavendish Laboratory, J.\ J.\ Thomson Avenue, Cambridge, CB3 0HE, UK}
\affiliation{Kavli Institute for Cosmology, Madingley Road, Cambridge, CB3 0HA, UK}
\author{M.\ P.\ Hobson}
\email{mph@mrao.cam.ac.uk}
\affiliation{Astrophysics Group, Cavendish Laboratory, J.\ J.\ Thomson Avenue, Cambridge, CB3 0HE, UK}
\author{W.\ J.\ Handley}
\email{wh260@cam.ac.uk}
\affiliation{Astrophysics Group, Cavendish Laboratory, J.\ J.\ Thomson Avenue, Cambridge, CB3 0HE, UK}
\affiliation{Kavli Institute for Cosmology, Madingley Road, Cambridge, CB3 0HA, UK}
\author{A.\ N.\ Lasenby}
\email{a.n.lasenby@mrao.cam.ac.uk}
\affiliation{Astrophysics Group, Cavendish Laboratory, J.\ J.\ Thomson Avenue, Cambridge, CB3 0HE, UK}
\affiliation{Kavli Institute for Cosmology, Madingley Road, Cambridge, CB3 0HA, UK}

\date{\today}

\begin{abstract}
We present a method to construct a continuous extension (otherwise known as dense output) for a numerical routine in the special case of the numerical solution being a scalar-valued function exhibiting rapid oscillations.
Such cases call for numerical routines that make use of the known global behaviour of the solution, one example being methods using asymptotic expansions to forecast the solution at each step of the independent variable. An example is \texttt{oscode}, numerical routine which uses the Wentzel--Kramers--Brillouin (WKB) approximation when the solution oscillates rapidly and otherwise behaves as a Runge--Kutta (RK) solver. 
Polynomial interpolation is not suitable for producing the solution at an arbitrary point mid-step, since efficient numerical methods based on the WKB approximation will step through multiple oscillations in a single step. Instead we construct the continuous solution by extending the numerical quadrature used in computing a WKB approximation of the solution with no additional evaluations of the differential equation or terms within, and provide an error estimate on this dense output. 
Finally, we draw attention to previous work on the continuous extension of Runge--Kutta formulae, and construct an extension to a RK method based on Gauss--Lobatto quadrature nodes, thus describing how to generate dense output from each of the methods underlying \texttt{oscode}.

\end{abstract}

\maketitle


\section{Introduction}

Ordinary differential equations with highly oscillatory solutions are ubiquitous in physics. Examples arise in quantum and celestial mechanics, in electrical circuits under alternating current, in the form of equations describing electromagnetic, pressure, or gravitational waves, and beyond. The efficient numerical solution of such equations requires specialised methods that can quickly traverse the rapid oscillations  as conventional methods available in commonly used scientific computing libraries (such as \texttt{scipy} \cite{scipy}, the \texttt{NAG} library \cite{naglib}, etc.) cannot do so. If highly oscillatory equations occur in the forward-modelling phase of Bayesian inference, their numerical solution can become the computational bottleneck. An example from the authors' field of research is inference from the Cosmic Microwave Background (CMB).  The analysis of extensions of the cosmological standard model (e.g.\ closed universe models \cite{LasenbyDoran,Handley-closed-1,Handley-closed-2}) is limited by the runtime of Boltzmann codes \cite{camb,class-1} that need to solve equations of the type described. It is therefore of great interest not only to develop numerical methods to solve such equations, but to make the methods robust and suitable for inclusion in mainstream numerical libraries. This requires the addition of advanced features such as dense output.

During the numerical solution of an ordinary differential equation (ODE) for an initial value problem, algorithms attempt to control the global error by adapting their stepsize - the spacing between values of the independent variable $x_i$ at which they naturally choose to evaluate the solution $y_i$ \cite{numerical_recipes}. The user might, however, wish to \emph{specify} the points at which the output is evaluated, for e.g. plotting the solution, event location, or treating discontinuities. The natural steps the numerical algorithm takes may be too large for such purposes. This problem is even more apparent if the user-specified tolerance is large, the numerical method is high-order or particularly efficient, or if the solution is smooth.
Artificially decreasing the steps (e.g. by integrating from one output point to the next) would be inefficient and increase computation time for a large number of outputs. 
One therefore relies on interpolation methods to generate an approximate solution mid-step, the process of which is referred to as dense output, and the generalisation of methods to yield dense output being termed a continuous extension of the method. Dense output should be produced with minimal computational overhead, i.e. with as few additional evaluations of the ODE as possible, and at a similar level of accuracy as that achieved at the natural steps. 

Interpolation of slowly changing functions in the context of dense output is well established \cite{RileyHobsonBence, numerical_recipes, AbramowitzStegun}, and often uses (piecewise) polynomials. There are instances, however, when polynomials are not applicable, one example being when the function to be interpolated undergoes several oscillations between two points of evaluation. There are several methods available to solve ODEs with highly oscillatory solutions efficiently (see, e.g. \cite{petzold, haddadin, Bamber2019, bremer}), which have in common that the global behaviour of the solution informs computation: they all exploit the prior knowledge that the solution is oscillatory. As a result, these algorithms may only `naturally' evaluate the solution every couple of oscillations, greatly reducing the number of steps taken, and requiring a different approach to computing dense output.

In this paper, we therefore present an approach that is based on identifying the slowly varying terms in the ODE from which the oscillatory solution can be constructed, and performing interpolation (based on polynomials) on these terms. More specifically, in this work we develop dense output for \texttt{oscode} \cite{Agocs2020}, a numerical method based on the Wentzel--Kramers--Brillouin (WKB) approximation \cite{BenderOrszag}, but the methodology shown is applicable to any solver using asymptotic expansions with non-oscillatory terms. Since \texttt{oscode} uses the WKB expansion to trace the solution in its oscillatory regimes but relies on a Runge--Kutta (RK) method otherwise, we gather results from the continuous extension of RK methods and adapt an existing method due to Shampine \cite{Shampine-practical} to compute dense output in the non-oscillatory regimes. 

The numerical method underlying \texttt{oscode} is reviewed briefly in \cref{oscode-overview}. \cref{dense-wkb} discusses dense output from steps when \texttt{oscode} uses the WKB expansion to forecast the solution, based on the continuous extension of Gaussian quadrature. We then derive a continuous extension of \texttt{oscode}'s custom RK method in \cref{dense-rk} before showing examples of dense output from both the WKB and RK regimes in \cref{examples} and concluding.

%
%
%
%

Throughout this paper we reserve the word integration for the process of solving an ODE, whereas the numerical evaluation of integrals will be referred to as quadrature.

\section{A WKB-based solver\label{oscode-overview}}

\texttt{oscode} is the implementation of a numerical method developed for the efficient solution of equations of the form
\begin{equation}\label{eq:eom}
    y'' + 2\gamma(x) y' + \omega^2(x) y = 0,
\end{equation}
where $y$ and $x$ are the dependent and independent variables, respectively.
This is the equation of motion of a one-dimensional, unforced, harmonic oscillator with a non-constant frequency (and damping term). Such equations are ubiquitous in physics and prove challenging to solve with conventional methods relying on a Taylor-series approximation to forecast the solution. This inspired a number of methods based on asymptotic expansions, \texttt{oscode} being one that uses the WKB expansion, which approximates the solution of (\ref{eq:eom}) analytically as
\begin{equation}\label{eq:wkb-form}
    y \sim A e^{\sum_{i=0}^{n} S_i(x) },  
\end{equation}
where the terms $S_i(x)$ can be derived recursively as
\begin{equation}\label{eq:wkb-si}
\begin{aligned}
    S_0 &= \pm i\int \omega dx, \\
    S_1 &= -\frac{1}{2}\ln \omega - \int \gamma dx, \\
    S_2 &= \pm i \int -\frac{1}{2}\frac{\gamma^2}{\omega} -\frac{1}{2}\frac{\gamma'}{\omega} + \frac{3}{8}\frac{\omega'^2}{\omega^3}-\frac{1}{4}\frac{\omega''}{\omega^2}dx, \\
    S_3 &= \frac{1}{4}\frac{\gamma^2}{\omega^2} + \frac{1}{4}\frac{\gamma'}{\omega^2} -  \frac{3}{16}\frac{\omega'^2}{\omega^4} + \frac{1}{8}\frac{\omega''}{\omega^3}, \\
    \ldots&, \\
    S'_i &= -\frac{1}{2S'_0}\left( S''_{i-1} + 2\gamma S'_{i-1} + \sum_{j=1}^{i-1} S'_{j}S'_{i-j}\right).
\end{aligned}    
\end{equation}
The above expansion approximates the real solution of (\ref{eq:eom}) in the limit of $\omega(x)$ varying on much longer scales than $y(x)$. It is a singular, perturbative, asymptotic expansion \cite{BenderOrszag}, which manifests itself in the sum  (\ref{eq:wkb-form})  usually being divergent. If the WKB approximation is valid in the region of interest, the successive terms in the series will each be much smaller then the previous, up until the smallest term $S_n$, at which the series should be truncated.

It is possible to embed the WKB approximation in a `stepping' procedure and apply it locally, rather than applying it to an entire oscillatory region. First, note that due to the $\pm$ sign in the $S_0$ term in (\ref{eq:wkb-si}), the WKB series yields two independent (approximate) solutions of the second order ODE (\ref{eq:eom}), thus all solutions can be obtained by the linear combination of
\begin{equation}
\begin{aligned}
    f_{+}(x) &= e^{S_0 + S_1 + S_2 + \ldots} \quad \mathrm{and} \\
    f_{-}(x) &= e^{-S_0 + S_1 - S_2 + \ldots}.
\end{aligned}
\end{equation}
If $y$ and $y'$ are known at $x$, then the solution can be forecast using the WKB series at a later point $x+h$ as
\begin{equation}
\begin{aligned}
    y(x+h) &= A_{+}f_{+}(x+h) + A_{-}f_{-}(x+h), \\
    y'(x+h) &= B_{+}f'_{+}(x+h) + B_{-}f'_{-}(x+h), \\
\end{aligned}
\end{equation}
where the $A_{\pm}$ and $B_{\pm}$ coefficients are functions of $y$, $f_{\pm}$, and their derivatives evaluated at the start of the step. Having different pairs of coefficients in the expressions for $y(x+h)$ and $y'(x+h)$ allows the phase of the WKB approximation to be reset at each step, rather than follow a single curve across steps \cite{Handley-RKWKB}.

The resulting algorithm is able to traverse many oscillations in a single step if the solution oscillates with a large, slowly changing frequency, at the cost of evaluating the numerical integrals and derivatives appearing in (\ref{eq:wkb-si}).
The key to producing dense output along such steps is thus computing dense output from the slowly varying terms within the $S_i(x)$ using the well-established methods available, and constructing the solution from them at any specified point.

It is to be noted that \texttt{oscode} is able to switch to a conventional method more suited for integrating non-oscillatory equations when necessary, its alternative method being a pair of RK formulas. In order for the algorithm to determine which of the two approximations to use, it requires two estimates of $y(x+h)$ to be evaluated at each step, one with each method. To maximise efficiency, both methods are based on the same set of evaluations of $\omega$ and $\gamma$, making the RK formula used a custom-made one rather than the highly optimal RK45 developed by Bogacki and Shampine \cite{BogackiShampine}, Dormand and Prince \cite{DormandPrince}, etc. Due to the customised nature of the RK formula used by \texttt{oscode} and for the sake of completeness, we summarise and apply the method of developing a continuous extension to RK formulas in \cref{dense-rk}.


\section{Dense output from the WKB expansion \label{dense-wkb}}

To construct the solution at an arbitrary point along the length of a step at the cost of just arithmetic operations (and no extra evaluations of terms in the ODE), we first review which numerical methods are used to construct the solution at the very end of the step, and which evaluations are available upon the completion of successful steps as a result.

\texttt{oscode} uses a WKB approximation expanded up to and including the $S_3$ term. If a WKB step runs from $x$ to $x+h$, the integrals in (\ref{eq:wkb-si}) will have those limits. For reasons stated in \cref{gauss-quad}, the integrals are computed using a form of Gaussian quadrature called Gauss--Lobatto rules. The quadrature method operates with $n$ evaluations of the integrand, two of which are always at the start and end of the step. To obtain an error estimate on the integrals, for each Gauss--Lobatto quadrature carried out with $n$ nodes there is one computed with $n-1$ nodes, the difference between the two giving the error estimate. As it will be discussed in \cref{gauss-quad}, Gaussian quadrature fits an interpolant to the integrand based on evaluations of the integrand at the $n$ nodes, hence it is possible to (1) evaluate this approximate integrand at any point, and (2) evaluate the integral itself at any point mid-step. This gives dense output on the $S_0$, $S_2$ terms, their derivatives, and the second term in $S_1$ together with its derivative. Dense output from the derivatives of the $S_i$ are required for constructing $y'$ at any point.

With the strategy for obtaining dense output from the numerical integrals discussed, we turn to the various derivatives of $\omega$ and $\gamma$ appearing in the WKB expansion. Since these are not available through the ODE directly, \texttt{oscode} uses the same trick RK methods are based upon: it combines evaluations of terms in the ODE ($\omega$, $\gamma$) at various values of $x < x_i < x+h$ such that when Taylor expanded, all terms lower than, and a maximal number of terms higher than the required derivative order vanish \cite{jordancalculus, Agocs2020}. This amounts to finding the coefficients for a finite difference equation which has fixed stencil points. As discussed, computation of the Gauss--Lobatto integrals $\int \omega dx$ and $\int \gamma dx$ requires $2n-3$ intermediate evaluations of $\omega$ and $\gamma$ per step, with \texttt{oscode} using $n=6$. These 9 evaluations in total can be used to obtain numerical derivatives sufficiently accurately at the nodes of Gauss--Lobatto integration, which can then be fit with the Gauss--Lobatto interpolant, and evaluated at arbitrary points.

\subsection{Gaussian quadrature\label{gauss-quad}}

The strategy behind Gaussian quadrature is to mimic the integrand $f(x)$ with a linear combination of orthogonal polynomials which have known integrals. The polynomials are chosen to best represent the integrand, different choices defining different quadrature rules. The linear combination $F(x)$ is fit to the integrand using a number of evaluations of the latter, such that they match at the abscissas $x=x_i$:
\begin{equation}
    \lim_{x\to x_i} F(x) = f(x).
\end{equation}
The integral then takes the form
\begin{equation}\label{eq:gauss-quad}
\begin{aligned}
    \tilde{I}(f) = \int_{a}^{b} F(s)ds &= \frac{b-a}{2}\int_{-1}^{1}F(s(x))dx \\
    &= \frac{b-a}{2}\sum_{i=1}^{n}{w_i f(x_i)}
\end{aligned}
\end{equation}
for a Gaussian method of $n$ nodes, with $w_i$ being weights of the method. Note that the integration limits have been shrunk down to $(-1,1)$ by a linear transformation of the independent variable,
\begin{equation}
    s = \frac{b-a}{2}x + \frac{b+a}{2}.
\end{equation}

The abscissas and weights can be chosen such that the order of the method, which in the context of quadrature means the degree for which all polynomials are integrated exactly by (\ref{eq:gauss-quad}), is much larger than $n$, e.g. $2n-1$ for Gauss--Legendre rules, and $2n-3$ for Gauss--Lobatto. Consequently, the remainder or error on the integral goes as a higher-order derivative of the integrand,
\begin{equation}
    R(f) \propto f^{(m)}(\xi), \quad -1<\xi<1,
\end{equation}
with $m=2n$ and $m=2n-2$ for Gauss--Legendre and Gauss--Lobatto rules, respectively. This makes Gaussian quadrature an especially attractive choice for performing the various numerical integrals appearing in the WKB approximation: if the WKB approximation is valid, the derivatives of the integrands involved will generally be small, therefore the applicability of the WKB approximation and Gaussian quadrature align well.

\subsection{Dense output from Gauss--Lobatto integration\label{gauss-lobatto}}

If the form of the interpolating polynomial $F(x)$ used by a quadrature rule is known, the numerical integral can be carried out until an arbitrary point within the integration limits, $a < c < b$, straightforwardly. 
Abscissas for Gauss--Lobatto integration are chosen to be the integration limits themselves ($x=\pm1$), and the roots of the polynomials
\begin{equation}
    P'_{n-1}(x),
\end{equation}
where $P_n(x)$ is the $n$th Legendre polynomial. The orthogonal polynomials associated with this quadrature rule then have to be 
\begin{equation}
    (1-x^2)P'_{n-1}(x),
\end{equation}
From this, the interpolation polynomial can be uniquely constructed as
\begin{eqnarray}
    F(x) &&= \frac{(1-x^2)P'_{n-1}(x)}{2P'_{n-1}(-1)(1+x)}f(-1) \nonumber\\
    &&+ \frac{(1-x^2)P'_{n-1}(x)}{2P'_{n-1}(1)(1-x)}f(1) \label{eq:integrand-glo}\\
    &&+ \sum_{i=2}^{n-2} \frac{(1-x^2)P'_{n-1}(x)}{P''_{n-1}(x_i)(1-x_i^2)(x-x_i)}f(x_i)\nonumber
\end{eqnarray}
Using l'H\^{o}pital's rule one can verify that this indeed reduces to $f(x_i)$ in the limit $x\to x_i$.

In the continuous extension of this method, the weights $w_i$ become position-dependent. If we wish to evaluate the solution at $x=c$, with $a < c < b$, we need to integrate (\ref{eq:integrand-glo}) up until that point,
\begin{equation}
\begin{aligned}
    \tilde{I}(f,c) = \int_{a}^{c}F(s)ds &= \frac{b-a}{2}\int_{-1}^{\tilde{c}}F(s(x))dx \\
    &= \frac{b-a}{2}\sum_{i=1}^{n} w_i(\tilde{c})f(x_i),
\end{aligned}
\end{equation}
where the weights are given by
\begin{gather}
    w_1(\tilde{c}) = \frac{1}{2P'_{n-1}(-1)}\int_{-1}^{\tilde{c}} (1-x)P'_{n-1}(x) dx, \\
    w_i(\tilde{c}) = \frac{1}{2P''_{n-1}(x_i)(1-x_i^2)}\int_{-1}^{\tilde{c}}\frac{(1-x^2)P'_{n-1}(x)}{x-x_i} dx, \\
    w_n(\tilde{c}) = \frac{1}{2P'_{n-1}(1)}\int_{-1}^{\tilde{c}} (1+x)P'_{n-1}(x), dx
\end{gather}
with $2 \geq i \geq n-1$.

\subsection{Error bound on dense output from Gauss--Lobatto integration\label{glo-error}}

\begin{figure*}
    \centering
    \includegraphics{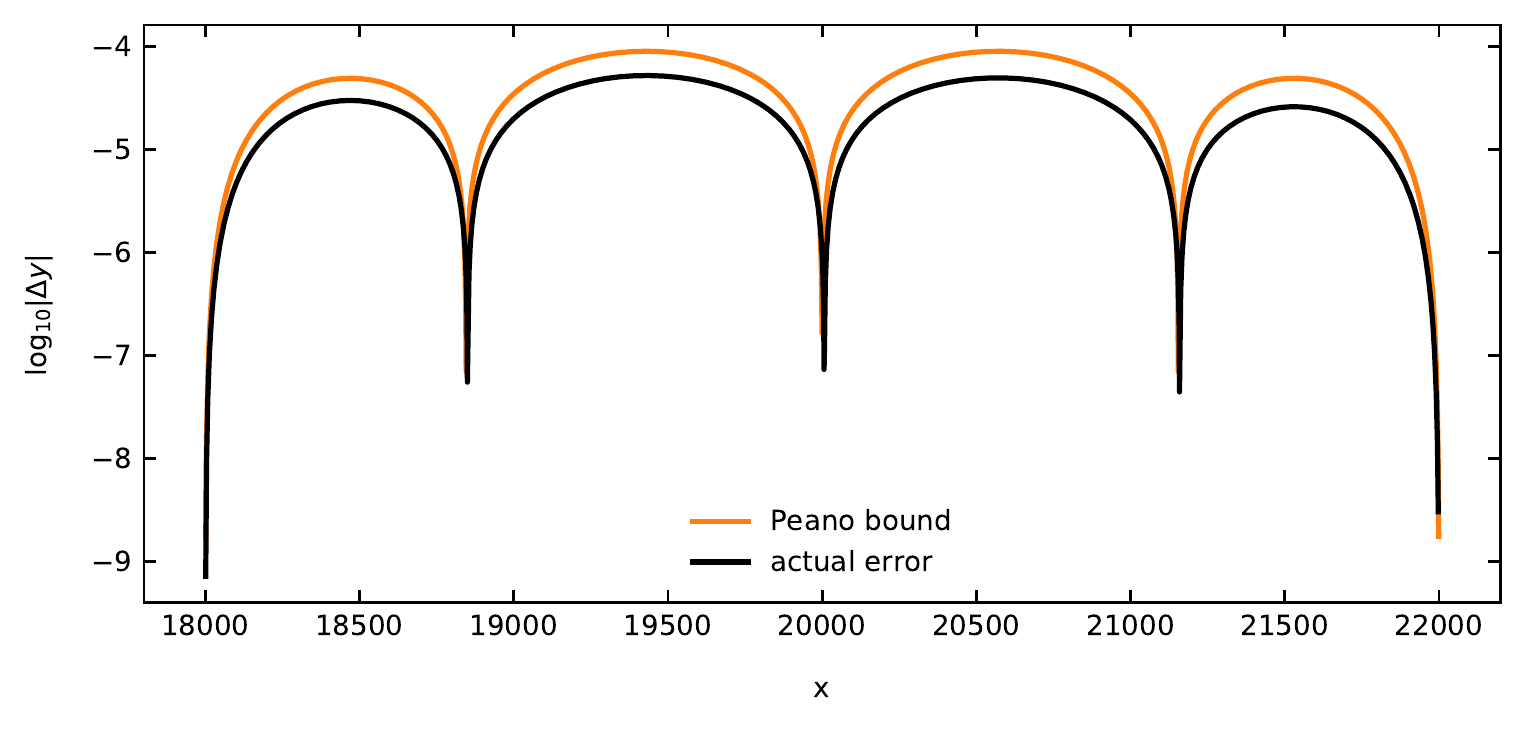}
    \caption{Base-10 logarithm of the residuals from the partial Gauss--Lobatto integration of $\omega =\sqrt{t}$ against the expected Peano error bound.}
\label{peano-2}
\end{figure*}

\begin{figure*}
    \centering
    \includegraphics{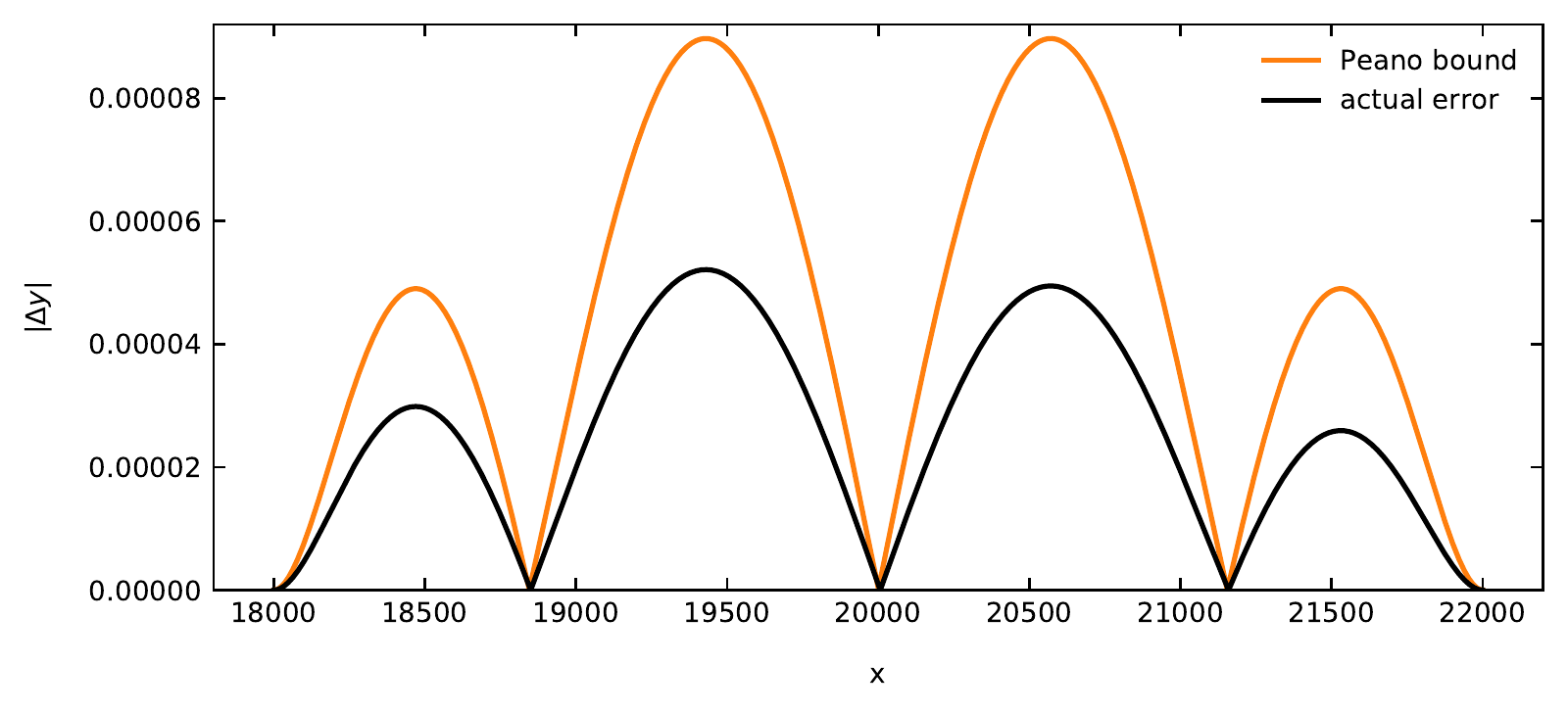}
    \caption{Actual residuals from the partial Gauss--Lobatto integration of $\omega =\sqrt{t}$ against the expected Peano error bound, shown with a linear $y$-axis..\label{peano-3}}
\end{figure*}

In \cref{gauss-quad} we claimed that the remainder of Gauss--Lobatto quadrature is proportional to $f^{(2n-2)}$, where $f(x)$ is the integrand and $n$ nodes are used to fit the interpolant to $f$. This is based on \emph{Peano's error representation} \cite{BulirschStoer}:
\begin{description}
    \item[Theorem] { Suppose $R(P) = 0$ holds for all polynomials $P \in \Pi_n$, i.e.\ every polynomial of degree less than or equal to $n$ is integrated exactly by the quadrature rules. Then for all functions $f \in C^{n+1}[a,b]$,
    \begin{equation}
        R(f) = \int_{a}^{b} f^{(n+1)}(t)K(t) dt,
    \end{equation}
    where
    \begin{equation}
        K(t) = \frac{1}{n!}R_{x}[ (x-t)^n_{+} ], \;\; (x-t)^n_{+} = \begin{cases} (x-t)^n \;\; \mathrm{if} \;\; x\geq t, \\ 0 \; \mathrm{if} x < t,\end{cases} 
    \end{equation}
    and 
    \begin{equation}
        R_x[(x-t)^n_{+}]
    \end{equation}
    is the remainder from the quadrature on $(x-t)^n_{+}$ when the latter is considered as a function of $x$.
    }
\end{description}
$K(t)$ is termed the Peano kernel of the remainder operator $R$. If the Peano kernel has constant sign on $[a,b]$ (which is the case for Gauss--Lobatto rules), it follows from the mean-value theorem of integral calculus that 
\begin{equation}\label{eq:peano-mvt}
    R(f) = f^{(n+1)}(\xi) \int_a^{b}K(t)dt \;\; \mathrm{for}\;\mathrm{some} \; \xi \in (a,b).
\end{equation}
The kernel $K(t)$ is the same for all integrands $f$ as long as the same quadrature rules are used. Therefore the integral $\int_a^{b}K(t)dt$ can be evaluated for any $f$, the simplest choice being the polynomial $f(x) = x^{n+1}$. Substituting this into (\ref{eq:peano-mvt}) and eliminating the kernel integral, one obtains
\begin{equation}\label{eq:remainder}
    R(f) = \frac{R(x^{n+1})}{(n+1)!}f^{(n+1)}(\xi) \;\; \mathrm{for}\;\mathrm{some} \; \xi \in (a,b).
\end{equation}
The elements of the above derivation do not depend on the integration limits $a,b$, and so the expression (\ref{eq:remainder}) for the remainder holds when applied to dense output, with one noteable difference.
Gauss--Lobatto quadrature, when performed `in full' on the interval $[a,b]$, integrates polynomials $P \in \Pi_{2n-3}$ exactly (the proof of which can also be found in \cite{BulirschStoer}), for which it is sometimes called a $2n-3$ order method. However, the same is not true if the upper integral limit is changed to $c < b$, while the nodes stay the same, as done in our computation of dense output. The proof of Gauss--Lobatto rules being of order $2n-3$ hinges on its basis polynomials being orthogonal on the interval $[a,b]$, which does not generally hold true on $[a,c]$. All one can say about the order of Gauss--Lobatto integration on the interval $[a,c]$ is that it is uniformly $n-1$, since the interpolant (\ref{eq:integrand-glo}) is a polynomial of degree $n-1$. Using (\ref{eq:remainder}), one can numerically (or otherwise) compute the Peano bound for the dense output from Gauss--Lobatto integration, which is seen in \cref{peano-2,peano-3}.

\section{Dense output from a Runge-Kutta method based on Gauss--Lobatto nodes\label{dense-rk}}

If it is suspected that the WKB approximation will not apply well in some regions of the integration range of an ODE, an alternative method is needed to efficiently traverse those regions. Runge--Kutta formulae are versatile and perform well in regions in which the solution is not oscillatory. An $n$th order Runge--Kutta method forecasts the solution $y_{n+1}=y(x_n+h)$ by Taylor-expanding around the beginning of the step $(x_n,y_n)$ and keeping terms in the Taylor series up to and including $h^ny^{(n)}/n!$, so the error on the step will be $\mathcal{O}(h^{n+1})$. The Taylor series is constructed using evaluations of the right-hand-side of the ODE,
\begin{equation}\label{eq:rk-ode}
    y' = G(x,y),
\end{equation}
as follows:
\begin{gather}
    k_0 = G(x_n,y_n), \nonumber \\
    k_i = G\left(x_n+c_ih, y_n + h\sum_{j=1}^{i-1} a_{ij}k_j\right), \; i=1..s, \label{eq:rk-butchert}\\
    y_{n+1} = y_n + h\sum_{i=1}^{s} b_i k_i. \nonumber  
\end{gather}
In the following section we will also make use of an evaluation of $G$ at the end of a successful step (which can be used in the next step in a first-same-as-last manner, and so is available `for free'),
\begin{equation}
    k_{s+1} = G(x_{n+1},y_{n+1}).
\end{equation}

\begin{table}[h]
\centering
\begin{tabular}{l|lllll}
0        &          &          &          &             &       \\
$c_2\quad$    & $\quad a_{21}\quad $ &          &          &             &       \\
$c_3\quad$    & $\quad a_{31}\quad $ & $\quad a_{32}\quad $ &          &             &       \\
$\vdots\quad$ & $\quad \vdots\quad $ & $\quad \vdots\quad $ & $\quad \ddots\quad $ & &       \\
$c_s\quad$    & $\quad a_{s1}\quad$ & $\quad a_{s2}\quad $ & $\quad \dotsm\quad $ & 
$\quad a_{s,s-1}\quad$ &       \\ \hline
         & $\quad b_1\quad $    & $\quad b_2\quad $    & $\quad \dotsm\quad $ & $\quad b_{s-1}\quad $   & $\quad b_s\quad $
\end{tabular}
\caption{Butcher tableau for an explicit Runge--Kutta method.\label{general_rk}}
\end{table}

The coefficients $a_{ij}$, $c_i$, and $b_i$ define the Runge--Kutta formula, and are often summarised in a Butcher tableau as in \cref{general_rk}. If $a_{ij} = 0$ for $j \leq i$, the formula is called explicit.
A formula that uses a number $s$ of $G$-evaluations is said to have $s$ stages.
For the formula to reproduce each term in the Taylor series up to the $n$th derivative, the coefficients have to satisfy a number of equations called order constraints. It can be shown that to satisfy the constraints at order $n > 4$, $s > n$ stages are required, with a minimum of $s=6$ for $n=5$, $s=7$ for $n=6$, and $s=9$ for $n=7$. While a higher order method allows for larger and fewer steps to keep the local error within the required tolerance, it is fruitless to increase the order to $n > 8$ due to the number of $G$-evaluations needed per step. For a method of a given $s$ and $n$ there will be leftover degrees of freedom once the order constraints are satisfied. These can be fixed by e.g. minimising the coefficient multiplying the $h^{n+1}$-term in the remainder, or to yield an $n-1$th order result using the same Butcher tableau entries (embedded methods).

In this work, we consider a 6-stage, 5th order explicit method that uses the nodes of the 6th order Gauss--Lobatto quadrature rules as its stages, i.e. the $c_i$ shall coincide with the $x_i$. This is so that when one computes a WKB approximation of the solution from $x=x_n$ to $x=x_n+h$, a Runge--Kutta step of the same size can be computed at the cost of a few arithmetic operations. The method however holds for any Runge--Kutta formula of the same $s$ and $n$, and can be generalised to others.

An approach to extend a Runge--Kutta formula developed by K. Horn \cite{horn-1983} is to take a hypothetical step from $x_n$ to $x_n+\sigma h$:
\begin{gather}
    k_0 = G(x_n,y_n), \nonumber \\
    k_i^{\ast} = G\left(x_n+c_i^{\ast}\sigma h, y_n + \sigma h\sum_{j=1}^{i-1} a_{ij}^{\ast}k_j^{\ast}\right), \; i=1..s^{\ast}, \label{eq:rk-cts}\\
    y_{n+1}^{\ast} = y_n + \sigma h\sum_{i=1}^{s} b_i^{\ast} k_i^{\ast}. \nonumber  
\end{gather}
The positions of $G$-evaluations can be made identical to those in the original Runge--Kutta formula if 
\begin{equation}
\begin{aligned}
    c_i^{\ast} &= \frac{1}{\sigma}c_i, \\
    a_{ij}^{\ast} &= \frac{1}{\sigma}a_{ij}.
\end{aligned}
\end{equation}
Depending on the order $n^{\ast}$ we require the new solution $y_{n+1}^{\ast}$ to be, it may be necessary to extend the Butcher tableau and add more stages to the formula, $s^{\ast} > s$. One can then derive a set of coefficients $b_i^{\ast}(\sigma)$ for each $\sigma$ that give the required order, and thus have a continuous extension. This approach, however, yields an answer such that
\begin{equation}
    \lim_{\sigma \to 1} y_{n+1}(\sigma) \neq y_{n+1},
\end{equation}
that is, the dense output would be discontinuous across steps. This is an undesirable property as ideally the dense output should be $C^1$ (continuously differentiable).

An alternative was proposed by Shampine in \cite{Shampine-practical} which uses Horn's interpolant to obtain a solution at one $\sigma$ at a similar order as that at achieved at the end of the step, then performs local polynomial interpolation based on this intermediate solution $y_{n+1}^{\ast}$, and information available at both ends of a step after a successful step. Horn has shown that for the Runge--Kutta--Fehlberg 4(5) formula, an intermediate solution at order 4 is available at $\sigma=0.6$ with $s^{\ast} = s$, i.e. for free. She did so by deriving the necessary order constraints for $b_i^{\ast}(\sigma)$ in a concise manner, which for $n^{\ast}=4$, $s^{\ast}=s=6$ are
\begin{equation}
\begin{aligned}
    \sum_{i=1}^{6} b^{\ast}_i c_i^j &= \frac{\sigma^j}{j+1} \quad \mathrm{for} \; i=0,1,2,3,\\
    \sum_{i=3}^{6} b_i^{\ast} a_{i2} &= 0. 
\end{aligned}
\end{equation}
Following Horn's procedure with the Butcher tableau entries derived in \cite{Agocs2020} and summarised in \cref{rk-glo-butcher}, we find that for this custom Runge--Kutta formula there is a free 4th order solution available at 
\begin{equation}
   \sigma = 0.58665886817 
\end{equation}
with the associated $b_i^{\ast}$ coefficients summarised in \cref{rk-glo-butcher-cts}. With this intermediate point, the following information is available: $y_n$, $\dot{y}_n$ through $k_0$, $y_{n+1}$ and $\dot{y}_{n+1}$ through $k_7$, and $y_{n+1}^{\ast}$. With these five constraints it is possible to carry out local quartic interpolation, meaning interpolation would be carried out for each successful step separately. This procedure, by relying on the solution and its derivative at both ends of a step, is ensured to provide a piecewise polynomial interpolant that is globally $C^1$.

    \begin{table}[h]
        \centering
        \footnotesize{
        \begin{tabular}{l|c}
        $c_1$ & 0  \\
        $c_2$ & $\frac{1}{2}\left(1-\sqrt{\frac{1}{3} + \frac{2\sqrt{7}}{21}}\right)$ \\
        $c_3$ & $\frac{1}{2}\left(1-\sqrt{\frac{1}{3} - \frac{2\sqrt{7}}{21}}\right)$ \\
        $c_4$ & $ \frac{1}{2}\left(1+\sqrt{\frac{1}{3} - \frac{2\sqrt{7}}{21}}\right)$ \\
        $c_5$ & $ \frac{1}{2}\left(1+\sqrt{\frac{1}{3} + \frac{2\sqrt{7}}{21}}\right) $ \\
        $c_6$ & $1$ \\ 
        $a_{21}$ & $ 0.117472338035267$ \\
        $a_{31}$ & $ -0.186247980065150 $ \\
        $a_{32}$ & $ 0.543632221824827 $ \\
        $a_{41}$ & $ -0.606430388550828  $ \\
        $a_{42}$ & $ 1  $ \\
        $a_{43}$ & $ 0.249046146791150  $ \\
        $a_{51}$ & $ 2.89935654001573 $ \\
        $a_{52}$ & $  -4.36852561156624  $ \\
        $a_{53}$ & $  2.13380671478631 $ \\
        $a_{54}$ & $  0.217890018728924 $ \\
        $a_{61}$ & $ 18.6799634999572 $\\
        $a_{62}$ & $ -28.8505778397313  $ \\
        $a_{63}$ & $ 10.7205340842092  $ \\
        $a_{64}$ & $ 1.41474175650804 $\\
        $a_{65}$ & $-0.964661500943270$ \\
        
        $b_1$ & $ 0.112755722735172$ \\
        $b_2$ & $0$ \\
        $b_3$ & $0.506557973265535$ \\
        $b_4$ & $ 0.0483004037699511$ \\
        $b_5$ & $0.378474956297846$ \\
        $b_6$ & $-0.0460890560685063$\\
        \end{tabular}
        }
        \caption{Butcher tableau for the 6-stage, 5${}^{\mathrm{th}}$ order Runge--Kutta method used by \texttt{oscode} in \cite{Agocs2020}. The points of function evaluations coincide with the abscissas of Gauss--Lobatto quadrature with~$n=6$. }
        \label{rk-glo-butcher}
    \end{table} 

    \begin{table}[h]
        \centering
        \footnotesize{
        \begin{tabular}{l|c}
        $b_1^{\ast}$ & $ 0.2089555395$ \\
        $b_2^{\ast}$ & $0$ \\
        $b_3^{\ast}$ & $0.7699501023$ \\
        $b_4^{\ast}$ & $0.009438629906$ \\
        $b_5^{\ast}$ & $-0.003746982422$ \\
        $b_6^{\ast}$ & $0.01540271068$\\
        \end{tabular}
        }
        \caption{Modified Butcher tableau entries for obtaining a 4th order estimate at $x=x_n+\sigma h$, with $\sigma = 0.58665886817$. The rest of the Butcher tableau entries match those in \cref{rk-glo-butcher}.}
        \label{rk-glo-butcher-cts}
    \end{table} 

In practice, the local interpolation can be performed efficiently as follows. 
Let the interpolating polynomial take the form 
\begin{equation}
y_{n+1} = y_n + a_1\sigma + a_2\sigma^2 + a_3\sigma^3 + a_4\sigma^4
\end{equation}
The constraints at the ends of a step and at the intermediate point $x_n + \sigma h$ can be written purely in terms of the $k_i$, the Butcher tableau coefficients, and the $a_1$--$a_4$:
\begin{equation}
\begin{aligned}
    k_0 &= a_1 \\
    \sum_{i=1}^{7} b_i k_i &= \sum_{i=1}^{4} a_i, \\
    k_7 &= \sum_{i=1}^{4} ia_i, \\
    \sigma \sum_{i=1}^{7} b_i^{\ast} k_i &= \sum_{i=1}^{4} \sigma^i a_i. \\
\end{aligned}
\end{equation}
From this one can extract the matrices $M$ and $S$ such that
\begin{equation}
\begin{aligned}
MQ^T &= SK, \\
 Q  &= \begin{bmatrix} a_1 & a_2 & a_3 & a_4 \end{bmatrix}, \\
 K^T  &= \begin{bmatrix} k_1 & k_2 & k_3 & k_4 & k_5 & k_6 & k_7 \end{bmatrix}.
\end{aligned}
\end{equation}
Note that if (\ref{eq:rk-ode}) is a vector equation, the $a_i$ and $k_i$ are promoted to column vectors and $Q$ and $K$ become matrices, but the notation used here still holds.
We can then extract a constant matrix $P$ from $Q$ via
\begin{equation}\label{eq:p-def}
    Q = K^T P,
\end{equation}
which lets us compute the dense output at the cost of the arithmetic operations underlying
\begin{equation}
    y = h Q Z,
\end{equation}
with
\begin{gather}
    Z = \begin{bmatrix} \sigma_1, \sigma_2, \ldots, \sigma_N \\ 
                             \sigma_1^2, \sigma_2^2, \ldots \sigma_N^2 \\
                             \sigma_1^3, \sigma_2^3, \ldots \sigma_N^3 \\
                             \sigma_1^4, \sigma_2^4, \ldots \sigma_N^4 
    \end{bmatrix},
\end{gather}
$N$ being the number of points we require output at.
$P$ can be pre-computed for each Runge--Kutta formula, and in our case is given in \cref{glo-rk-P}. This procedure is used in open-source scientific computing libraries such as \texttt{scipy} \cite{scipy}.

\begin{table}[h]
    \centering
    \footnotesize{
    \begin{tabular}{c|c|c|c}
    $1$ & $ -2.48711376 $ & $ 2.42525041$ & $-0.82538093  $ \\
    $ 0 $ & $0 $  & $0 $ & $0 $\\
    $0 $ & $ 3.78546138$  & $ -5.54469086$ & $2.26578746 $\\
    $0 $ & $-0.27734213 $  & $ 0.74788587$ & $ -0.42224334$\\
    $0 $ & $-2.94848704 $  & $ 7.41087391 $ & $-4.08391191 $\\
    $0 $ & $ 0.50817346 $ & $-1.20070313 $ & $  0.64644062$\\
    $0 $ & $ 1.4193081$ & $ -3.8386162 $ & $ 2.4193081 $\\
    \end{tabular}
    }
    \caption{Pre-computed $P$-matrix for a 5th order, 6-stage Runge--Kutta formula based on 6 Gauss--Lobatto nodes, as defined in (\ref{eq:p-def}).}
    \label{glo-rk-P}
\end{table}


\begin{figure*}
    \centering
    \includegraphics{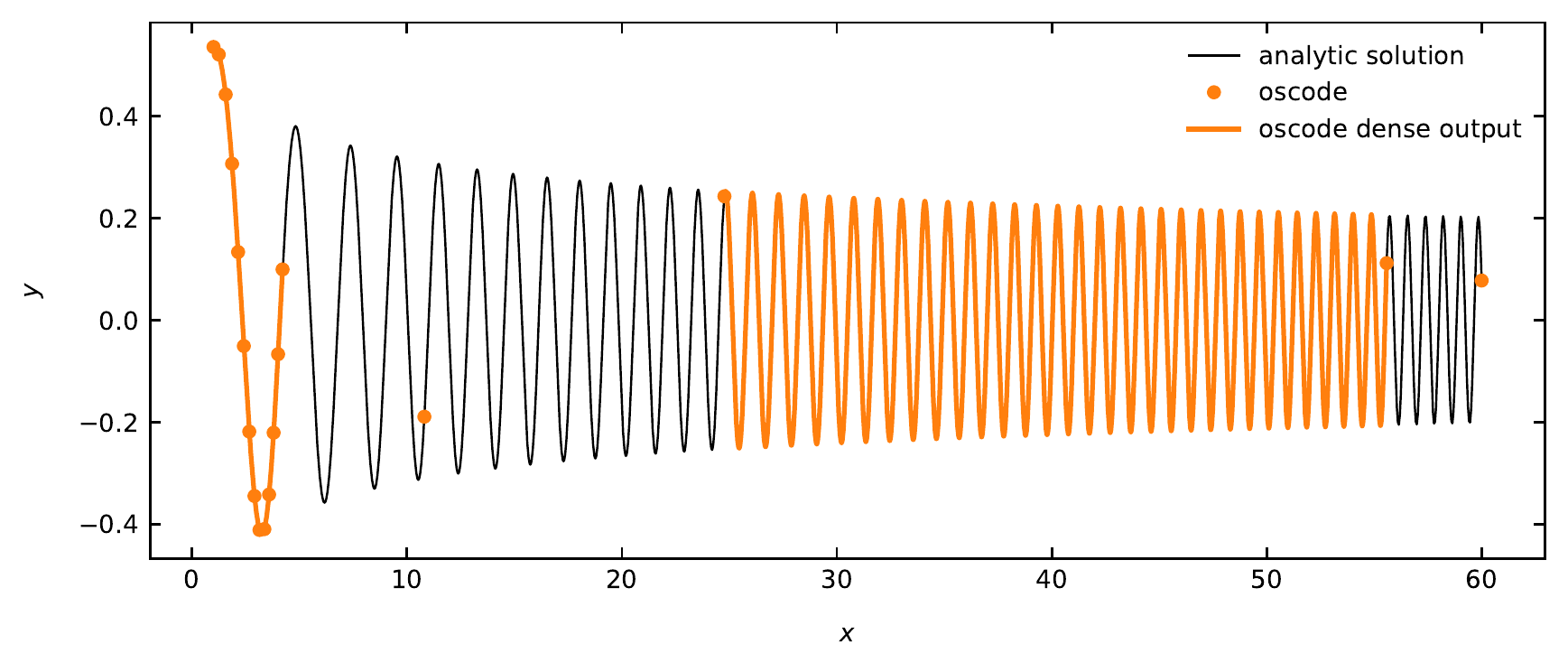}
    \caption{Dense output from \texttt{oscode} solving the Airy equation. As the frequency becomes larger but more slowly-changing, the method switches from using the RK method to the WKB approximation and the distance between natural steps (orange dots) of the algorithm increases. The first segment of solid, orange line on top of the analytic solution (black line) denotes dense output from the RK method used by \texttt{oscode}, with the second orange segment showing dense output throughout a WKB step.  }
    \label{airy_demo}
\end{figure*}

\begin{figure*}
    \centering
    \includegraphics{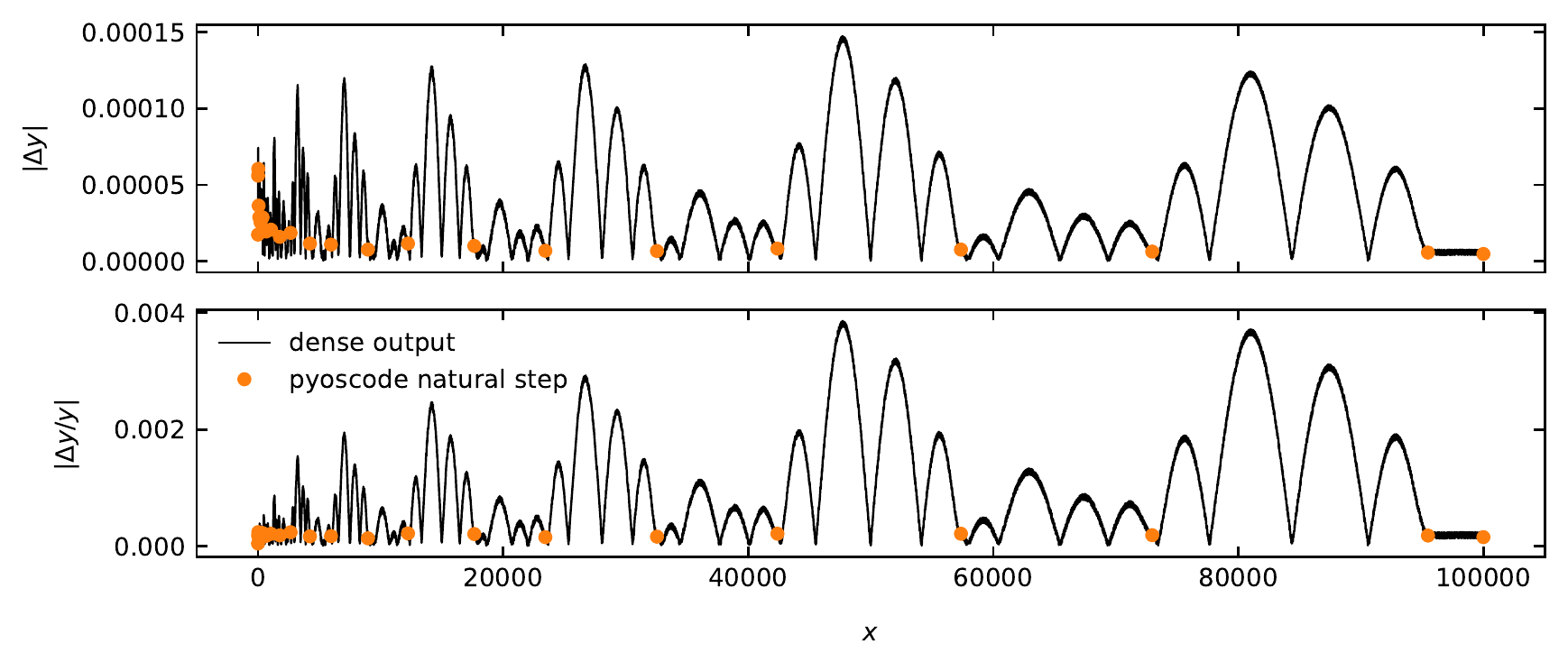}
    \caption{Absolute (top) and relative (bottom) residuals from dense output of the Airy equation, relative to the analytic solution. Note that the error at the ends of natural steps (orange dots) are much smaller than throughout the steps, due to Gauss--Lobatto integration being much higher ($2n-3$) order at the end of a step than mid-way ($n-1$). }
    \label{airy_res}
\end{figure*}

\begin{figure*}
    \centering
    \includegraphics{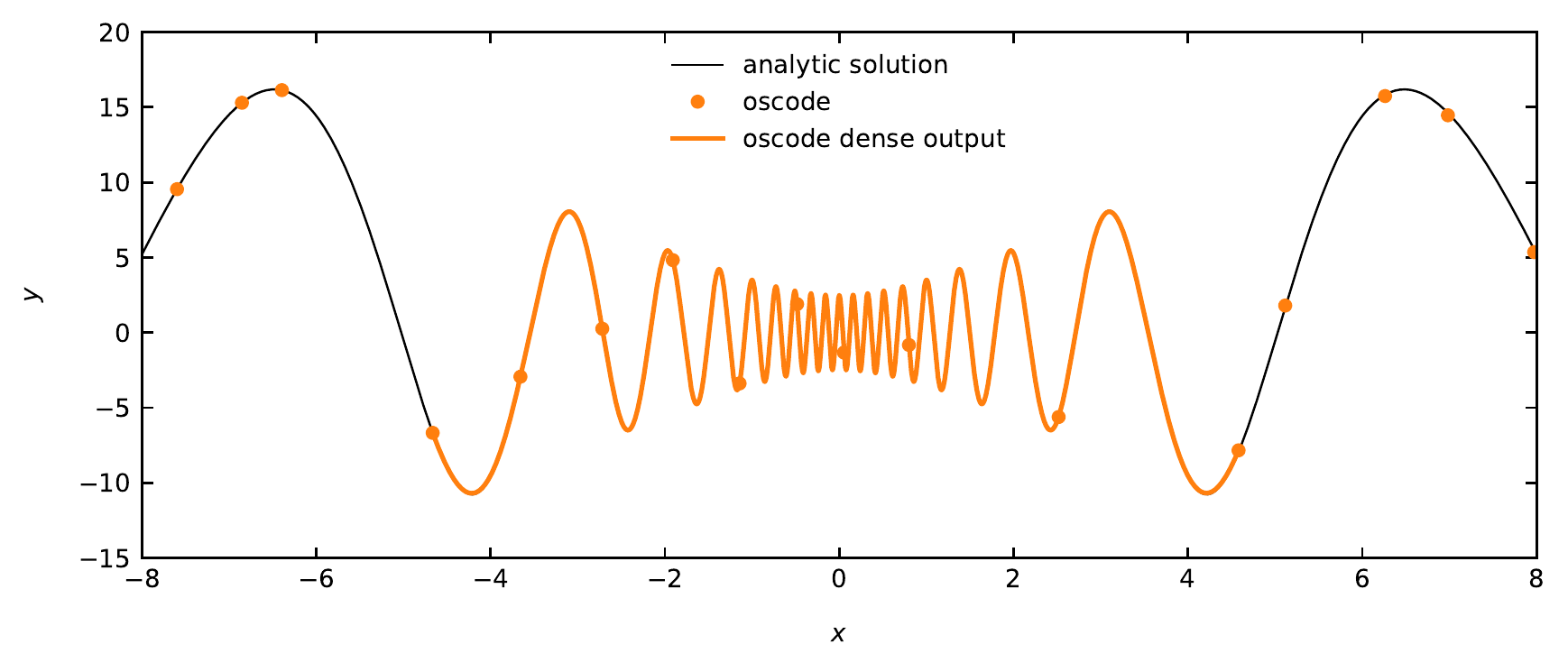}
    \caption{Dense output from \texttt{oscode} solving (\ref{eq:burst}), an equation exhibiting a burst of oscillations. }
    \label{burst_demo}
\end{figure*}



\section{Examples\label{examples}}

We now show a few examples of dense output from the WKB expansion and a RK method as implemented in \texttt{oscode}. 

The Airy equation,
\begin{equation}
    y'' + xy = 0,
\end{equation}
describes an oscillator with a frequency $\omega(x) = \sqrt{x}$ that rises with increasing $x$, but at a decreasing rate. As $x$ increases, therefore, it becomes more favourable for \texttt{oscode} to use the WKB approximation to forecast the solution, and it switches over to do so from using a RK method at around $x \sim 5$, as shown in \cref{airy_demo}. The figure shows dense output from both the initial RK and the late WKB phase on top of the analytic solution. The error on the dense output from the numerical solution of the Airy equation (for a longer integration range) is shown in \cref{airy_res}, exhibiting a pattern similar to that seen in \cref{peano-2}. The similarity is due to the leading term in the numerical error coming from the $S_0 = \int \omega(x)dx $ term of the WKB expansion. At large values of $x$ the WKB approximation is valid, and so successive terms decrease rapidly in the expansion.
\cref{burst_demo} shows dense output from the numerical solution of the equation
\begin{equation}\label{eq:burst}
    y'' + \frac{n^2-1}{(1+x^2)^2}y = 0
\end{equation}
with $n=40$. The parameter $n$ governs the number of oscillations the solution exhibits around $x=0$. When \texttt{oscode} solves equation, it uses the WKB approximation around $x=0$ and uses RK otherwise, switching in a symmetrical manner. The error properties and performance of \texttt{oscode} are explored in \cite{Agocs2020} using the example of (\ref{eq:burst}), but we note that due to no extra evaluations of the terms in this ODE being made during the computation of dense output, the latter does not increase the overall computing time significantly.

\section{Conclusions\label{conclusions}}

Dense output (evaluation of the numerical solution of an ODE at user-specified points) cannot always be constructed by polynomial interpolation between the natural steps of a numerical algorithm. One example is the efficient solution of ODEs with highly oscillatory solutions, for in this case the algorithm may traverse many oscillations in a single step. Such equations are extremely common in physics, and often form computational bottlenecks in e.g.\ the forward-modelling phase of Bayesian inference, when tackled with conventional (Runge--Kutta-like) methods.

Out of the methods available to efficiently solve highly oscillatory ODEs, \texttt{oscode} uses the WKB approximation to forecast the solution many wavelengths ahead if the characteristic frequency of oscillations changes on a much longer timescale than the solution itself, and otherwise behaves as a RK solver. In this work we proposed procedures to generate dense output from each of the methods underlying \texttt{oscode}. 

In a region where the one-dimensional solution of an ODE oscillates and the WKB approximation is valid, we propose to perform interpolation with known methods on the slowly-changing frequency ($\omega$) and damping ($\gamma$) terms in the ODE, and construct the solution using the WKB approximation. 
The numerical integrals of $\omega$ and $\gamma$ appearing in the WKB expansion can be computed efficiently and to high accuracy with Gaussian quadrature methods. We summarised dense output from one method of the Gaussian family, Gauss--Lobatto quadrature, and derived an error bound for the output.

In regions where the RK method is more appropriate, we reviewed existing techniques to obtain a continuous extension of RK methods, and demonstrated them on the example of a RK method based on nodes of 6-point Gauss--Lobatto quadrature.

\begin{acknowledgments}
FJA thanks Nils Sch\"{o}eneberg for fruitful discussions. She was supported by the Science and Technology Facilities Council. WJH was supported by a Gonville \& Caius college research fellowship.
\end{acknowledgments}



\nocite{*}

\bibliography{apssamp}

\providecommand{\noopsort}[1]{}\providecommand{\singleletter}[1]{#1}%
\begin{thebibliography}{29}%
\makeatletter
\providecommand \@ifxundefined [1]{%
 \@ifx{#1\undefined}
}%
\providecommand \@ifnum [1]{%
 \ifnum #1\expandafter \@firstoftwo
 \else \expandafter \@secondoftwo
 \fi
}%
\providecommand \@ifx [1]{%
 \ifx #1\expandafter \@firstoftwo
 \else \expandafter \@secondoftwo
 \fi
}%
\providecommand \natexlab [1]{#1}%
\providecommand \enquote  [1]{``#1''}%
\providecommand \bibnamefont  [1]{#1}%
\providecommand \bibfnamefont [1]{#1}%
\providecommand \citenamefont [1]{#1}%
\providecommand \href@noop [0]{\@secondoftwo}%
\providecommand \href [0]{\begingroup \@sanitize@url \@href}%
\providecommand \@href[1]{\@@startlink{#1}\@@href}%
\providecommand \@@href[1]{\endgroup#1\@@endlink}%
\providecommand \@sanitize@url [0]{\catcode `\\12\catcode `\$12\catcode
  `\&12\catcode `\#12\catcode `\^12\catcode `\_12\catcode `\%12\relax}%
\providecommand \@@startlink[1]{}%
\providecommand \@@endlink[0]{}%
\providecommand \url  [0]{\begingroup\@sanitize@url \@url }%
\providecommand \@url [1]{\endgroup\@href {#1}{\urlprefix }}%
\providecommand \urlprefix  [0]{URL }%
\providecommand \Eprint [0]{\href }%
\providecommand \doibase [0]{http://dx.doi.org/}%
\providecommand \selectlanguage [0]{\@gobble}%
\providecommand \bibinfo  [0]{\@secondoftwo}%
\providecommand \bibfield  [0]{\@secondoftwo}%
\providecommand \translation [1]{[#1]}%
\providecommand \BibitemOpen [0]{}%
\providecommand \bibitemStop [0]{}%
\providecommand \bibitemNoStop [0]{.\EOS\space}%
\providecommand \EOS [0]{\spacefactor3000\relax}%
\providecommand \BibitemShut  [1]{\csname bibitem#1\endcsname}%
\let\auto@bib@innerbib\@empty
\bibitem [{\citenamefont {{Virtanen}}\ \emph {et~al.}(2020)\citenamefont
  {{Virtanen}}, \citenamefont {{Gommers}}, \citenamefont {{Oliphant}},
  \citenamefont {{Haberland}}, \citenamefont {{Reddy}}, \citenamefont
  {{Cournapeau}}, \citenamefont {{Burovski}}, \citenamefont {{Peterson}},
  \citenamefont {{Weckesser}}, \citenamefont {{Bright}}, \citenamefont {{van
  der Walt}}, \citenamefont {{Brett}}, \citenamefont {{Wilson}}, \citenamefont
  {{Jarrod Millman}}, \citenamefont {{Mayorov}}, \citenamefont {{Nelson}},
  \citenamefont {{Jones}}, \citenamefont {{Kern}}, \citenamefont {{Larson}},
  \citenamefont {{Carey}}, \citenamefont {{Polat}}, \citenamefont {{Feng}},
  \citenamefont {{Moore}}, \citenamefont {{Vand erPlas}}, \citenamefont
  {{Laxalde}}, \citenamefont {{Perktold}}, \citenamefont {{Cimrman}},
  \citenamefont {{Henriksen}}, \citenamefont {{Quintero}}, \citenamefont
  {{Harris}}, \citenamefont {{Archibald}}, \citenamefont {{Ribeiro}},
  \citenamefont {{Pedregosa}}, \citenamefont {{van Mulbregt}},\ and\
  \citenamefont {{Contributors}}}]{scipy}%
  \BibitemOpen
  \bibfield  {author} {\bibinfo {author} {\bibfnamefont {P.}~\bibnamefont
  {{Virtanen}}}, \bibinfo {author} {\bibfnamefont {R.}~\bibnamefont
  {{Gommers}}}, \bibinfo {author} {\bibfnamefont {T.~E.}\ \bibnamefont
  {{Oliphant}}}, \bibinfo {author} {\bibfnamefont {M.}~\bibnamefont
  {{Haberland}}}, \bibinfo {author} {\bibfnamefont {T.}~\bibnamefont
  {{Reddy}}}, \bibinfo {author} {\bibfnamefont {D.}~\bibnamefont
  {{Cournapeau}}}, \bibinfo {author} {\bibfnamefont {E.}~\bibnamefont
  {{Burovski}}}, \bibinfo {author} {\bibfnamefont {P.}~\bibnamefont
  {{Peterson}}}, \bibinfo {author} {\bibfnamefont {W.}~\bibnamefont
  {{Weckesser}}}, \bibinfo {author} {\bibfnamefont {J.}~\bibnamefont
  {{Bright}}}, \bibinfo {author} {\bibfnamefont {S.~J.}\ \bibnamefont {{van der
  Walt}}}, \bibinfo {author} {\bibfnamefont {M.}~\bibnamefont {{Brett}}},
  \bibinfo {author} {\bibfnamefont {J.}~\bibnamefont {{Wilson}}}, \bibinfo
  {author} {\bibfnamefont {K.}~\bibnamefont {{Jarrod Millman}}}, \bibinfo
  {author} {\bibfnamefont {N.}~\bibnamefont {{Mayorov}}}, \bibinfo {author}
  {\bibfnamefont {A.~R.~J.}\ \bibnamefont {{Nelson}}}, \bibinfo {author}
  {\bibfnamefont {E.}~\bibnamefont {{Jones}}}, \bibinfo {author} {\bibfnamefont
  {R.}~\bibnamefont {{Kern}}}, \bibinfo {author} {\bibfnamefont
  {E.}~\bibnamefont {{Larson}}}, \bibinfo {author} {\bibfnamefont
  {C.}~\bibnamefont {{Carey}}}, \bibinfo {author} {\bibfnamefont
  {{\.I}.}~\bibnamefont {{Polat}}}, \bibinfo {author} {\bibfnamefont
  {Y.}~\bibnamefont {{Feng}}}, \bibinfo {author} {\bibfnamefont {E.~W.}\
  \bibnamefont {{Moore}}}, \bibinfo {author} {\bibfnamefont {J.}~\bibnamefont
  {{Vand erPlas}}}, \bibinfo {author} {\bibfnamefont {D.}~\bibnamefont
  {{Laxalde}}}, \bibinfo {author} {\bibfnamefont {J.}~\bibnamefont
  {{Perktold}}}, \bibinfo {author} {\bibfnamefont {R.}~\bibnamefont
  {{Cimrman}}}, \bibinfo {author} {\bibfnamefont {I.}~\bibnamefont
  {{Henriksen}}}, \bibinfo {author} {\bibfnamefont {E.~A.}\ \bibnamefont
  {{Quintero}}}, \bibinfo {author} {\bibfnamefont {C.~R.}\ \bibnamefont
  {{Harris}}}, \bibinfo {author} {\bibfnamefont {A.~M.}\ \bibnamefont
  {{Archibald}}}, \bibinfo {author} {\bibfnamefont {A.~H.}\ \bibnamefont
  {{Ribeiro}}}, \bibinfo {author} {\bibfnamefont {F.}~\bibnamefont
  {{Pedregosa}}}, \bibinfo {author} {\bibfnamefont {P.}~\bibnamefont {{van
  Mulbregt}}}, \ and\ \bibinfo {author} {\bibfnamefont {S.~.~.}\ \bibnamefont
  {{Contributors}}},\ }\href {\doibase
  https://doi.org/10.1038/s41592-019-0686-2} {\bibfield  {journal} {\bibinfo
  {journal} {Nature Methods}\ }\textbf {\bibinfo {volume} {17}},\ \bibinfo
  {pages} {261} (\bibinfo {year} {2020})}\BibitemShut {NoStop}%
\bibitem [{\citenamefont {{The Numerical Algorithms Group (NAG), Oxford, United
  Kingdom}}()}]{naglib}%
  \BibitemOpen
  \bibfield  {author} {\bibinfo {author} {\bibnamefont {{The Numerical
  Algorithms Group (NAG), Oxford, United Kingdom}}},\ }\href@noop {} {\enquote
  {\bibinfo {title} {{The NAG Library}},}\ }\bibinfo {howpublished}
  {\url{http://www.nag.com/}}\BibitemShut {NoStop}%
\bibitem [{\citenamefont {{Lasenby}}\ and\ \citenamefont
  {{Doran}}(2005)}]{LasenbyDoran}%
  \BibitemOpen
  \bibfield  {author} {\bibinfo {author} {\bibfnamefont {A.}~\bibnamefont
  {{Lasenby}}}\ and\ \bibinfo {author} {\bibfnamefont {C.}~\bibnamefont
  {{Doran}}},\ }\href {\doibase 10.1103/PhysRevD.71.063502} {\bibfield
  {journal} {\bibinfo  {journal} {Physical Review D}\ }\textbf {\bibinfo
  {volume} {71}},\ \bibinfo {eid} {063502} (\bibinfo {year} {2005})},\ \Eprint
  {http://arxiv.org/abs/astro-ph/0307311} {astro-ph/0307311} \BibitemShut
  {NoStop}%
\bibitem [{\citenamefont {Handley}(2019{\natexlab{a}})}]{Handley-closed-1}%
  \BibitemOpen
  \bibfield  {author} {\bibinfo {author} {\bibfnamefont {W.}~\bibnamefont
  {Handley}},\ }\href@noop {} {\enquote {\bibinfo {title} {Curvature tension:
  evidence for a closed universe},}\ } (\bibinfo {year} {2019}{\natexlab{a}}),\
  \Eprint {http://arxiv.org/abs/1908.09139} {arXiv:1908.09139 [astro-ph.CO]}
  \BibitemShut {NoStop}%
\bibitem [{\citenamefont {Handley}(2019{\natexlab{b}})}]{Handley-closed-2}%
  \BibitemOpen
  \bibfield  {author} {\bibinfo {author} {\bibfnamefont {W.}~\bibnamefont
  {Handley}},\ }\href {\doibase 10.1103/physrevd.100.123517} {\bibfield
  {journal} {\bibinfo  {journal} {Physical Review D}\ }\textbf {\bibinfo
  {volume} {100}} (\bibinfo {year} {2019}{\natexlab{b}}),\
  10.1103/physrevd.100.123517}\BibitemShut {NoStop}%
\bibitem [{\citenamefont {Lewis}\ \emph {et~al.}(2000)\citenamefont {Lewis},
  \citenamefont {Challinor},\ and\ \citenamefont {Lasenby}}]{camb}%
  \BibitemOpen
  \bibfield  {author} {\bibinfo {author} {\bibfnamefont {A.}~\bibnamefont
  {Lewis}}, \bibinfo {author} {\bibfnamefont {A.}~\bibnamefont {Challinor}}, \
  and\ \bibinfo {author} {\bibfnamefont {A.}~\bibnamefont {Lasenby}},\ }\href
  {\doibase 10.1086/309179} {\bibfield  {journal} {\bibinfo  {journal} {The
  Astrophysical Journal}\ }\textbf {\bibinfo {volume} {538}},\ \bibinfo {pages}
  {473–476} (\bibinfo {year} {2000})}\BibitemShut {NoStop}%
\bibitem [{\citenamefont {{Lesgourgues}}(2011)}]{class-1}%
  \BibitemOpen
  \bibfield  {author} {\bibinfo {author} {\bibfnamefont {J.}~\bibnamefont
  {{Lesgourgues}}},\ }\href@noop {} {\bibfield  {journal} {\bibinfo  {journal}
  {arXiv e-prints}\ ,\ \bibinfo {eid} {arXiv:1104.2932}} (\bibinfo {year}
  {2011})},\ \Eprint {http://arxiv.org/abs/1104.2932} {arXiv:1104.2932
  [astro-ph.IM]} \BibitemShut {NoStop}%
\bibitem [{\citenamefont {Press}\ \emph {et~al.}(2007)\citenamefont {Press},
  \citenamefont {Teukolsky}, \citenamefont {Vetterling},\ and\ \citenamefont
  {Flannery}}]{numerical_recipes}%
  \BibitemOpen
  \bibfield  {author} {\bibinfo {author} {\bibfnamefont {W.~H.}\ \bibnamefont
  {Press}}, \bibinfo {author} {\bibfnamefont {S.~A.}\ \bibnamefont
  {Teukolsky}}, \bibinfo {author} {\bibfnamefont {W.~T.}\ \bibnamefont
  {Vetterling}}, \ and\ \bibinfo {author} {\bibfnamefont {B.~P.}\ \bibnamefont
  {Flannery}},\ }\href@noop {} {\emph {\bibinfo {title} {Numerical Recipes 3rd
  Edition: The Art of Scientific Computing}}},\ \bibinfo {edition} {3rd}\ ed.\
  (\bibinfo  {publisher} {Cambridge University Press},\ \bibinfo {address}
  {USA},\ \bibinfo {year} {2007})\BibitemShut {NoStop}%
\bibitem [{\citenamefont {Riley}\ \emph {et~al.}(2006)\citenamefont {Riley},
  \citenamefont {Hobson},\ and\ \citenamefont {Bence}}]{RileyHobsonBence}%
  \BibitemOpen
  \bibfield  {author} {\bibinfo {author} {\bibfnamefont {K.~F.}\ \bibnamefont
  {Riley}}, \bibinfo {author} {\bibfnamefont {M.~P.}\ \bibnamefont {Hobson}}, \
  and\ \bibinfo {author} {\bibfnamefont {S.~J.}\ \bibnamefont {Bence}},\
  }\href@noop {} {{\selectlanguage {english}\emph {\bibinfo {title}
  {Mathematical methods for physics and engineering}}}},\ \bibinfo {edition}
  {3rd}\ ed.\ (\bibinfo  {publisher} {Cambridge University Press},\ \bibinfo
  {address} {Cambridge},\ \bibinfo {year} {2006})\BibitemShut {NoStop}%
\bibitem [{\citenamefont {Abramowitz}(1965)}]{AbramowitzStegun}%
  \BibitemOpen
  \bibfield  {author} {\bibinfo {author} {\bibfnamefont {M.}~\bibnamefont
  {Abramowitz}},\ }\href@noop {} {{\selectlanguage {english}\emph {\bibinfo
  {title} {Handbook of mathematical functions, with formulas, graphs, and
  mathematical tables.}}}}\ (\bibinfo  {publisher} {Dover},\ \bibinfo {address}
  {New York},\ \bibinfo {year} {1965})\BibitemShut {NoStop}%
\bibitem [{\citenamefont {Petzold}(1981)}]{petzold}%
  \BibitemOpen
  \bibfield  {author} {\bibinfo {author} {\bibfnamefont {L.~R.}\ \bibnamefont
  {Petzold}},\ }\href {http://www.jstor.org/stable/2156866} {\bibfield
  {journal} {\bibinfo  {journal} {SIAM Journal on Numerical Analysis}\ }\textbf
  {\bibinfo {volume} {18}},\ \bibinfo {pages} {455} (\bibinfo {year}
  {1981})}\BibitemShut {NoStop}%
\bibitem [{\citenamefont {Haddadin}\ and\ \citenamefont
  {Handley}(2018)}]{haddadin}%
  \BibitemOpen
  \bibfield  {author} {\bibinfo {author} {\bibfnamefont {W.~I.~J.}\
  \bibnamefont {Haddadin}}\ and\ \bibinfo {author} {\bibfnamefont {W.~J.}\
  \bibnamefont {Handley}},\ }\href@noop {} {\enquote {\bibinfo {title} {Rapid
  numerical solutions for the mukhanov-sazaki equation},}\ } (\bibinfo {year}
  {2018}),\ \Eprint {http://arxiv.org/abs/1809.11095} {arXiv:1809.11095
  [astro-ph.CO]} \BibitemShut {NoStop}%
\bibitem [{\citenamefont {{Bamber}}\ and\ \citenamefont
  {{Handley}}(2019)}]{Bamber2019}%
  \BibitemOpen
  \bibfield  {author} {\bibinfo {author} {\bibfnamefont {J.}~\bibnamefont
  {{Bamber}}}\ and\ \bibinfo {author} {\bibfnamefont {W.}~\bibnamefont
  {{Handley}}},\ }\href@noop {} {\bibfield  {journal} {\bibinfo  {journal}
  {{arXiv e-prints }}\ } (\bibinfo {year} {2019})},\ \Eprint
  {http://arxiv.org/abs/1907.11638} {arXiv:1907.11638 [physics.comp-ph]}
  \BibitemShut {NoStop}%
\bibitem [{\citenamefont {Bremer}(2018)}]{bremer}%
  \BibitemOpen
  \bibfield  {author} {\bibinfo {author} {\bibfnamefont {J.}~\bibnamefont
  {Bremer}},\ }\href {\doibase https://doi.org/10.1016/j.acha.2016.05.002}
  {\bibfield  {journal} {\bibinfo  {journal} {Applied and Computational
  Harmonic Analysis}\ }\textbf {\bibinfo {volume} {44}},\ \bibinfo {pages} {312
  } (\bibinfo {year} {2018})}\BibitemShut {NoStop}%
\bibitem [{\citenamefont {Agocs}\ \emph {et~al.}(2020)\citenamefont {Agocs},
  \citenamefont {Handley}, \citenamefont {Lasenby},\ and\ \citenamefont
  {Hobson}}]{Agocs2020}%
  \BibitemOpen
  \bibfield  {author} {\bibinfo {author} {\bibfnamefont {F.~J.}\ \bibnamefont
  {Agocs}}, \bibinfo {author} {\bibfnamefont {W.~J.}\ \bibnamefont {Handley}},
  \bibinfo {author} {\bibfnamefont {A.~N.}\ \bibnamefont {Lasenby}}, \ and\
  \bibinfo {author} {\bibfnamefont {M.~P.}\ \bibnamefont {Hobson}},\ }\href
  {\doibase 10.1103/PhysRevResearch.2.013030} {\bibfield  {journal} {\bibinfo
  {journal} {Phys. Rev. Research}\ }\textbf {\bibinfo {volume} {2}},\ \bibinfo
  {pages} {013030} (\bibinfo {year} {2020})}\BibitemShut {NoStop}%
\bibitem [{\citenamefont {Bender}\ and\ \citenamefont
  {Orszag}(1999)}]{BenderOrszag}%
  \BibitemOpen
  \bibfield  {author} {\bibinfo {author} {\bibfnamefont {C.~M.}\ \bibnamefont
  {Bender}}\ and\ \bibinfo {author} {\bibfnamefont {S.~A.}\ \bibnamefont
  {Orszag}},\ }\href@noop {} {{\selectlanguage {english}\emph {\bibinfo {title}
  {Advanced mathematical methods for scientists and engineers. I, Asymptotic
  methods and perturbation theory}}}}\ (\bibinfo  {publisher} {Springer},\
  \bibinfo {address} {New York, N.Y.},\ \bibinfo {year} {1999})\BibitemShut
  {NoStop}%
\bibitem [{\citenamefont {Shampine}(1986)}]{Shampine-practical}%
  \BibitemOpen
  \bibfield  {author} {\bibinfo {author} {\bibfnamefont {L.~F.}\ \bibnamefont
  {Shampine}},\ }\href {http://www.jstor.org/stable/2008219} {\bibfield
  {journal} {\bibinfo  {journal} {Mathematics of Computation}\ }\textbf
  {\bibinfo {volume} {46}},\ \bibinfo {pages} {135} (\bibinfo {year}
  {1986})}\BibitemShut {NoStop}%
\bibitem [{\citenamefont {{Handley}}\ \emph {et~al.}(2016)\citenamefont
  {{Handley}}, \citenamefont {{Lasenby}},\ and\ \citenamefont
  {{Hobson}}}]{Handley-RKWKB}%
  \BibitemOpen
  \bibfield  {author} {\bibinfo {author} {\bibfnamefont {W.~J.}\ \bibnamefont
  {{Handley}}}, \bibinfo {author} {\bibfnamefont {A.~N.}\ \bibnamefont
  {{Lasenby}}}, \ and\ \bibinfo {author} {\bibfnamefont {M.~P.}\ \bibnamefont
  {{Hobson}}},\ }\href@noop {} {\bibfield  {journal} {\bibinfo  {journal}
  {{arXiv e-prints }}\ } (\bibinfo {year} {2016})},\ \Eprint
  {http://arxiv.org/abs/1612.02288} {arXiv:1612.02288 [physics.comp-ph]}
  \BibitemShut {NoStop}%
\bibitem [{\citenamefont {Bogacki}\ and\ \citenamefont
  {Shampine}(1996)}]{BogackiShampine}%
  \BibitemOpen
  \bibfield  {author} {\bibinfo {author} {\bibfnamefont {P.}~\bibnamefont
  {Bogacki}}\ and\ \bibinfo {author} {\bibfnamefont {L.}~\bibnamefont
  {Shampine}},\ }\href {\doibase https://doi.org/10.1016/0898-1221(96)00141-1}
  {\bibfield  {journal} {\bibinfo  {journal} {Computers \& Mathematics with
  Applications}\ }\textbf {\bibinfo {volume} {32}},\ \bibinfo {pages} {15 }
  (\bibinfo {year} {1996})}\BibitemShut {NoStop}%
\bibitem [{\citenamefont {Dormand}\ and\ \citenamefont
  {Prince}(1980)}]{DormandPrince}%
  \BibitemOpen
  \bibfield  {author} {\bibinfo {author} {\bibfnamefont {J.}~\bibnamefont
  {Dormand}}\ and\ \bibinfo {author} {\bibfnamefont {P.}~\bibnamefont
  {Prince}},\ }\href {\doibase https://doi.org/10.1016/0771-050X(80)90013-3}
  {\bibfield  {journal} {\bibinfo  {journal} {Journal of Computational and
  Applied Mathematics}\ }\textbf {\bibinfo {volume} {6}},\ \bibinfo {pages} {19
  } (\bibinfo {year} {1980})}\BibitemShut {NoStop}%
\bibitem [{\citenamefont {Jordan}\ and\ \citenamefont
  {Jord{\'a}n}(1965)}]{jordancalculus}%
  \BibitemOpen
  \bibfield  {author} {\bibinfo {author} {\bibfnamefont {C.}~\bibnamefont
  {Jordan}}\ and\ \bibinfo {author} {\bibfnamefont {K.}~\bibnamefont
  {Jord{\'a}n}},\ }\href@noop {} {\emph {\bibinfo {title} {Calculus of finite
  differences}}},\ Vol.~\bibinfo {volume} {33}\ (\bibinfo  {publisher}
  {American Mathematical Soc.},\ \bibinfo {year} {1965})\BibitemShut {NoStop}%
\bibitem [{\citenamefont {Stoer}\ and\ \citenamefont
  {Roland}(1983)}]{BulirschStoer}%
  \BibitemOpen
  \bibfield  {author} {\bibinfo {author} {\bibfnamefont {J.}~\bibnamefont
  {Stoer}}\ and\ \bibinfo {author} {\bibfnamefont {B.}~\bibnamefont {Roland}},\
  }\href@noop {} {{\selectlanguage {english}\emph {\bibinfo {title}
  {Introduction to numerical analysis}}}}\ (\bibinfo  {publisher}
  {Springer-Verlag},\ \bibinfo {address} {New York},\ \bibinfo {year}
  {1983})\BibitemShut {NoStop}%
\bibitem [{\citenamefont {Horn}(1983)}]{horn-1983}%
  \BibitemOpen
  \bibfield  {author} {\bibinfo {author} {\bibfnamefont {M.~K.}\ \bibnamefont
  {Horn}},\ }\href {http://www.jstor.org/stable/2157271} {\bibfield  {journal}
  {\bibinfo  {journal} {SIAM Journal on Numerical Analysis}\ }\textbf {\bibinfo
  {volume} {20}},\ \bibinfo {pages} {558} (\bibinfo {year} {1983})}\BibitemShut
  {NoStop}%
\bibitem [{\citenamefont {Teukolsky}(2015)}]{Teukolsky2015}%
  \BibitemOpen
  \bibfield  {author} {\bibinfo {author} {\bibfnamefont {S.~A.}\ \bibnamefont
  {Teukolsky}},\ }\href {\doibase 10.1016/j.jcp.2014.12.012} {\bibfield
  {journal} {\bibinfo  {journal} {Journal of Computational Physics}\ }\textbf
  {\bibinfo {volume} {283}},\ \bibinfo {pages} {408–413} (\bibinfo {year}
  {2015})}\BibitemShut {NoStop}%
\bibitem [{\citenamefont {Hunter}\ and\ \citenamefont
  {Nikolov}(2000)}]{Hunter2000}%
  \BibitemOpen
  \bibfield  {author} {\bibinfo {author} {\bibfnamefont {D.}~\bibnamefont
  {Hunter}}\ and\ \bibinfo {author} {\bibfnamefont {G.}~\bibnamefont
  {Nikolov}},\ }\href@noop {} {\bibfield  {journal} {\bibinfo  {journal}
  {Mathematics of computation}\ }\textbf {\bibinfo {volume} {69}},\ \bibinfo
  {pages} {269} (\bibinfo {year} {2000})}\BibitemShut {NoStop}%
\bibitem [{\citenamefont {Hunter}(1995)}]{Hunter1995}%
  \BibitemOpen
  \bibfield  {author} {\bibinfo {author} {\bibfnamefont {D.~B.}\ \bibnamefont
  {Hunter}},\ }\href {\doibase 10.1007/BF01732979} {\bibfield  {journal}
  {\bibinfo  {journal} {BIT Numerical Mathematics}\ }\textbf {\bibinfo {volume}
  {35}},\ \bibinfo {pages} {64} (\bibinfo {year} {1995})}\BibitemShut {NoStop}%
\bibitem [{\citenamefont {Schira}(1996)}]{Schira1996}%
  \BibitemOpen
  \bibfield  {author} {\bibinfo {author} {\bibfnamefont {T.}~\bibnamefont
  {Schira}},\ }\href {\doibase https://doi.org/10.1016/S0377-0427(96)00100-8}
  {\bibfield  {journal} {\bibinfo  {journal} {Journal of Computational and
  Applied Mathematics}\ }\textbf {\bibinfo {volume} {76}},\ \bibinfo {pages}
  {171 } (\bibinfo {year} {1996})}\BibitemShut {NoStop}%
\bibitem [{\citenamefont {Schira}(1997)}]{Schira1996-symmetric}%
  \BibitemOpen
  \bibfield  {author} {\bibinfo {author} {\bibfnamefont {T.}~\bibnamefont
  {Schira}},\ }\href {http://www.jstor.org/stable/2153655} {\bibfield
  {journal} {\bibinfo  {journal} {Mathematics of Computation}\ }\textbf
  {\bibinfo {volume} {66}},\ \bibinfo {pages} {297} (\bibinfo {year}
  {1997})}\BibitemShut {NoStop}%
\bibitem [{\citenamefont {Stenger}(1966)}]{Stenger1966}%
  \BibitemOpen
  \bibfield  {author} {\bibinfo {author} {\bibfnamefont {F.}~\bibnamefont
  {Stenger}},\ }\href {\doibase 10.1007/BF02163184} {\bibfield  {journal}
  {\bibinfo  {journal} {Numerische Mathematik}\ }\textbf {\bibinfo {volume}
  {8}},\ \bibinfo {pages} {150} (\bibinfo {year} {1966})}\BibitemShut {NoStop}%
\end{thebibliography}%

\end{document}